\documentclass[]{JHEP3} 

\let\ifpdf\relax
\usepackage{ifpdf}

\usepackage{epsfig,multicol,bbm}
\usepackage[sort&compress,numbers]{natbib}

\usepackage{graphicx}

\newcommand{\beq}{\begin{equation}}
\newcommand{\eeq}{\end{equation}}
\newcommand{\beqa}{\begin{eqnarray}}
\newcommand{\eeqa}{\end{eqnarray}}
\newcommand{\nn}{\nonumber}
\newcommand{\brem}{bremsstrahlung }

\title{Improved partonic event generators at lepton colliders}

\author{Walter T. Giele\\
Fermilab, P.O.Box 500, Batavia, IL 60510, USA\\
E-mail: \email{giele@fnal.gov}
}

\preprint{FERMILAB-PUB-15-122-T}
\abstract{
A method is detailed for the phase space integration of 
$\gamma^*\rightarrow$ multi-jets applicable to parton level Monte
Carlo's at any order in perturbation theory. Other non-jet objects,
massless or massive, can be included in the phase space generation.
We correlate the \brem events in a manner that integrates out
all partonic configurations leading to a fixed jet configuration,
thereby improving convergence. This also allows the method to extend
infra-red safety to the fully differential multi-jet cross section.
}

\keywords{QCD}

\begin{document} 

\section{Introduction}

When calculating differential cross sections involving jets the majority of
effort and ingenuity is invested in evaluating the matrix
elements. The phase space integration and definitions of proper
observables are usually afterthoughts. 
For Leading Order (LO) parton level Monte Carlo's many methods have
been proposed to improve the behavior of multi-jet event
generation. These methods focus on the properties of the LO
amplitude~\cite{Kleiss:1994qy,Draggiotis:2000gm,vanHameren:2000aj,vanHameren:2002tc,
 Gleisberg:2008fv,vanHameren:2010gg,Papadopoulos:2000tt}.
Here we focus on event generation for parton level event generators
beyond LO.  At higher order partons are clustered, resulting in jets composed of two or more partons. This
internal configuration of the jet is not purview to the observer
within the context of a fixed order calculation and
needs to be averaged, resulting in a perturbative calculable and reliable result.   

We will exploit this necessity by constructing a phase space generator
which explicitly integrates out all partonic configurations which
give rise to the same multi-jet finals state for a particular jet algorithm.
In ref.~\cite{Giele:2011tm} the basic concepts of such an approach to
phase space integration was developed and a
proof of existence was given. These techniques were refined and
modified in ref.~\cite{Campbell:2012cz} for application
to the Matrix Element Method (MEM), extending the concept of MEM to Next-to-Leading
Order (NLO)~\cite{Campbell:2013hz}.  Finally, a first look at how to
implement these correlated phase space generators for the calculation
of NLO cross sections were highlighted in
ref.~\cite{Campbell:2013uha}. In this paper we develop the method
fully for parton event generators at lepton colliders, ready to be
applied to e.g. fully differential Next-to-Next-to-Leading Order (NNLO) 3-jet cross sections~\cite{Ridder:2014wza}.

The standard manner in which phase space event generators work in
Monte Carlo's calculating higher order correction to jet cross sections
is to generate the events with different parton multiplicities
completely uncorrelated. For example, to calculate the NNLO
3-jet cross section at a leptonic collider one first generates many
3-parton events and bins the observable calculated from the 3 momenta,
weighted by the value of the sum of the regulated 3-parton amplitudes.
Once completed, another run is initiated generating many 4-parton
events for which the observable  is calculated and binned with the
weight of the regulated 4-parton amplitude. Finally, the 5-parton
events are generated and the observable binned using the regulated 5-parton amplitude.
In this the integration of the observable over the jet phase space plays a crucial role to achieve
infra-red safety.  

Yet, from the viewpoint
of the observed jet final state, these different parton multiplicity events
have a high degree of correlation. By applying a jet algorithm, the hadronic final state is
simplified into a final state of jet objects, each jet having a momentum. Perturbative QCD
should be able to calculate these jet cross sections, i.e. the
correlations between the jet momenta, order by order in perturbation
theory provided the jets are opaque. That is, the jets are averaged over all
hadronic configurations leading to that particular jet configuration.
As a consequence the fully differential jet cross section should be calculable
given an appropriate infra-red safe jet algorithm. For the NNLO 3-jet
example this will significantly alter the manner in which the 3-parton, 4-parton
and 5-parton final states are generated. First of all, the generator
calculates the fully differential 3-jet cross section
$d\,\sigma_3/d\,p_1d\,p_2d\,p_3$ where $\{p_i\}$ are the jet axis
momenta. Observables can be calculated by integrating the jet observable over the jet phase space. 
Note that nowhere it is required to average over the jet phase space to obtain an
infra-red safe answer, nor are there any constraints  on the type of
jet observable one can look at.
For the event generator this means that the starting point is a given fixed 3-jet
configuration (i.e. the jet momenta $\{p_i\}$ are given). This requires
only a single evaluation of the 3-parton amplitudes. From this
starting point we generate 4-parton events which reconstruct back
to the initiating 3-jet momenta using the jet algorithm. Because the jets are fixed the 4-parton
amplitude weights can simply be added to the 3-parton amplitude
weight. Due to the recursive nature of sequential clustering in
jet algorithm we can repeat the above argument for the 5-parton
contribution: given a single 4-parton event we can generate many
5-parton events such that applying a single step in the jet
algorithm leads back to the initiating 4-parton event. It is clear
that this will lead to a highly correlated event generator and the
event weight of the fully differential 3-jet cross section is a series
expansion in the strong coupling constant at the scale of the jet resolution. 

Current jet algorithms do not extend the concept of infra-red safety
to the fully differential jet cross sections. This is solely due to the
clustering phase in the algorithm, i.e. how to combine two momenta to form a new
one. The new momentum formed from the momenta of the respective clustered
particles is simply the sum of these momenta. For some
algorithms this new momentum is redefined to make it a
massless momentum (see
ref. ~\cite{Bethke:1991wk} for an overview of these schemes). 
As a result applying the jet algorithm on e.g. a 4-parton final state to form a
3-jet final state will never overlap with the 3-jet final state
generated from the 3-parton final state due to jet mass and/or non-momentum conservation. 

As a consequence we have to adjust the cluster phase of the jet algorithms
in order to obtain infra-red safe fully differential jet cross sections.
Note that one can still apply any ordinary jet
algorithm using the above described correlated phase space generator. 
However the now theoretical jet algorithm used internally by the
generator to reorganize the correlated phase space generation does
not match the applied jet algorithm exactly.
The price to pay is that one must define observables and
average this jet observable by integrating over the
full jet phase space so that an infra-red safe prediction can be made.
Still, the generation of the correlated multi-parton final
states could greatly benefit the convergence of the predictions of the
observable made by the Monte Carlo. 

In section 2 we will construct the required forward brancher which
converts an $n$-particle phase space into a $(n+1)$-particle phase
space, invertible using the cluster algorithm. In section 3
this Forward Branching Phase Space (FBPS) generator is used to
construct the correlated event generators needed for calculating fully
differential multi-jet cross sections at any order in perturbation theory.
We conclude in section 3 by summarizing the results and outline the
next steps in further developing the FBPS method.

\section{Constructing the forward branching phase space generator}

The first step in the construction of the FBPS is to define a proper
clustering algorithm. As already explained in the introduction, we need a
clustering algorithm which maps a massless $n$-particle phase space
onto a massless $(n-1)$-particles phase.
\footnote{Note that it is straightforward to add other, massive or
  massless, momenta as long as they to not participate in the jet
  clustering. For the remainder of the paper we ignore this in order
  to simplify the notation and discussion.} 

Current jet algorithms simply combine two momenta by adding the
4-vectors. Take as an example the decay of a heavy particle with
momentum $Q$ into four massless particles, clustering momentum $p_3$
and momentum $p_4$
\beq\label{clusterbasic}
Q=\hat p_1+\hat p_2+\hat p_3+\hat p_4=\hat p_1+\hat p_2+\hat p_{34}\ .
\eeq
This obviously lead to a 3-particle phase space where one of the
momenta is massive. Some clustering algorithms rescale the
energy or momentum such that the clustered particle is massless (see
ref.~\cite{Bethke:1991wk}). However such a scheme would change 
$Q$ through momentum conservation.

The way to modify the clustering is to use a recoil momentum. By
rescaling at least one of the other non-clustered particles one can
make the clustered particle massless (this is in essence a $3\rightarrow
2$ clustering as is commonly used to construct the subtraction terms
needed to regulate various amplitudes in higher order
calculations~\cite{Catani:1996vz}). 
Explicitly, eq.~\ref{clusterbasic} becomes
\beq\label{clustermod}
Q=\hat p_1+\hat p_2+\hat p_3+\hat p_4=\hat p_1+\hat p_2+\hat 
p_{34}=(1-\beta) ( \hat p_1+\hat p_2)+(\hat p_{34}+\beta ( \hat
p_1+\hat p_2))=p_1+p_2+p_3\ ,
\eeq 
where we have used the sum of all other particles as a recoil momentum
so to avoid further complicating the algorithm on how to select a
recoil particle. The new momenta are given by $p_1=(1-\beta)\hat p_1$,
$p_2=(1-\beta)\hat p_2$ and $p_3=\hat p_{34}+\beta(\hat p_1+\hat p_2)$. 
The branching scale variable $\beta$ can now be chosen such that the
clustered particle is massless
\beqa
p_3^2&=&(\hat p_{34}+\beta\hat p_{12})^2=\beta^2\hat
s_{12}+2\beta\,\hat p_{34}\cdot p_{12}+\hat
p_{34}^2=0\nn\\
\Rightarrow\beta&=&\frac{\left(\hat p_{34}\cdot\hat
    p_{12}\right)-\sqrt{\left(\hat p_{34}\cdot\hat
      p_{12}\right)^2-\hat s_{12}\hat p_{34}^2}}{\hat s_{12}}\ ,
\eeqa
were $p_{ij}=p_i+p_j$ and $s_{ij}=p_{ij}^2$. With this simple augmentation of the
cluster algorithm we have what we need to construct the cluster
invertible FBPS generators.

It is convenient to introduce here the inverse of the clustering
kinematics, i.e. the branching kinematics. Introducing the branching
scaling variable $\alpha$ we have
\beq
Q=p_1+p_2+p_3=(1+\alpha)(p_1+p_2)+(p_3-\alpha (p_1+p_2))=\hat p_1+\hat
p_2+\hat p_{34}=\hat p_1+\hat p_2+\hat p_3+\hat p_4\ ,
\eeq
where $\alpha$ is given by the quadratic equation
\beq
\hat s_{34}=\hat p_{34}^2=(p_3-\alpha (p_1+p_2))^2=\alpha^2
s_{12}-2\alpha\, p_3\cdot (p_1+p_2)\ .
\eeq
The branching and merging scale variables are related by
\beq
1-\beta=\frac{1}{1+\alpha}\Rightarrow \beta=\frac{\alpha}{1+\alpha}\ ,
\eeq
and give the relations to be used later
\beq
\beta(\hat p_1+\hat p_2)=\beta\hat p_{12}=\beta(1+\alpha)p_{12}=\alpha
p_{12}=\alpha(p_1+p_2) ;\ d\,\beta=\frac{d\,\alpha}{(1+\alpha)^2}\ .
\eeq

Having augmented the clustering algorithm we can now start making the
FBPS generator needed to generate the massless $(n+1)$-parton phase space
from the massless $n$-parton phase space subject to the clustering
constraint of eq.~\ref{clustermod}.

The $n$-particle fully differential cross section is given by
\beq
\frac{d\,\sigma}{d\,p_1\cdots
  d\,p_n}=\left(\frac{(2\pi)^4}{2\sqrt{Q^2}}\right)\times
d\,\Phi_n(Q;p_1,\ldots,p_n)\ .
\eeq
The flat phase space is given by
\beq
d\,\Phi_n(Q;p_1,\ldots,p_n)=\prod_{i=1}^n \frac{d^4
p_i}{(2\pi)^3} \delta^+(p_i^2)\,\delta^4(Q-p_1-\cdots-p_n)\ .
\eeq

We will derive the FBPS generating a massless 4-particle phase space
from a massless 3-particle phase space. Afterward we will generalize
this to $n$-particle phase spaces.
The first step in constructing the FBPS generator is simply generating
the massless 4-particle phase space from a 3-particle phase space with
one massive particle
\beq
d\,\Phi_4(Q;\hat p_1\hat p_2\hat p_3\hat p_4)=
(2\pi)^3d \hat s_{34}\ \Phi_3(Q;\hat p_1\hat p_2\hat
p_{34})\times\Phi_2(\hat p_{34};\hat p_1\hat p_2)\ ,
\eeq
where $\hat p_{34}^2=\hat s_{34}$.

The next step is to implement the rescaling of momentum $\hat p_{34}$
using the recoil momentum $\hat p_{12}$. Specifically, starting from
the massive phase space
\beq\label{PSmassive}
d\Phi_3(Q;\hat p_1\hat p_2\hat p_{34})=\frac{d^4\hat
  p_1}{(2\pi)^3}\frac{d^4\hat p_2}{(2\pi)^3}\frac{d^4\hat p_{34}}{(2\pi)^3}
\delta^+(\hat p_1^2)\delta^+(\hat p_2^2)\delta^+(\hat p_{34}^2-\hat
s_{34})\delta^4(Q-\hat p_1-\hat p_2-\hat p_{34})\ ,
\eeq
we have to derive the Jacobean $J$ generated by the change of the
integration momenta
\beqa
d\Phi_3(Q;\hat p_1\hat p_2\hat p_{34})&=&J\times
d\Phi_3(Q;p_1p_2p_3)\\
&=&J\times \frac{d^4
  p_1}{(2\pi)^3}\frac{d^4 p_2}{(2\pi)^3}\frac{d^4 p_3}{(2\pi)^3}
\delta^+(p_1^2)\delta^+(p_2^2)\delta^+(p_3^2)\delta^4(Q-p_1-p_2-p_3)\ .
\nn
\eeqa

To calculate the Jacobian $J$ we need to rewrite the integrals over
the momenta $\hat p_1$,$\hat p_2$ and $\hat p_{34}$ into integrals
over momenta $p_1$, $p_2$ and $p_3$. To do this we mathematically
express the clustering algorithm as a decomposition of unity
\beqa
1&=&\hat s_{12} (\beta_+-\beta_-)\int d\,\beta\,d\,p_1\,d\,p_2\,
d\,p_3\delta((1-\beta)\hat p_1-p_1)\delta((1-\beta)\hat
p_2-p_2)\delta(p_3^2)\nn\\
&\times&\delta(p_3-(\hat p_{34}+\beta\hat p_{12}))\ ,
\eeqa
where $\beta_\pm$ are given by solving equation $(\hat
p_{34}+\beta\hat p_{12})^2=0$:
\beq
\hat s_{12}(\beta_+-\beta_-)=2\sqrt{(\hat p_{34}\cdot\hat
  p_{12})^2-\hat s_{34}\hat s_{12}}\ .
\eeq
The above identity is expressed in terms of the cluster scale variable
$\beta$ and cluster kinematics.  We rewrite this equation in terms of the branching scale variable
$\alpha$ and branching kinematics, giving us
\beqa
1&=&2\int d\,\alpha\,d\,p_1\,d\,p_2\,d\,p_3
(1+\alpha)^7\sqrt{(\hat p_{34}\cdot
  p_{12})^2-\hat s_{34} s_{12}}
\nn\\
&\times&\delta(\hat p_1-(1+\alpha) p_1) \delta(\hat p_2-(1+\alpha)
p_2)\delta(p_3^2)\delta(\hat p_{34}-(p_3-\alpha\, p_{12}))\ .
\eeqa
Multiplying this decomposition of one to eq.~\ref{PSmassive} and integrating the
momenta $\hat p_1$, $\hat p_2$ and $\hat p_{34}$ over the appropriate 
$\delta$-functions one obtains
\beqa
d\Phi_3(Q;\hat p_1\hat p_2\hat p_{34})&=&\int d\,\alpha
(1+\alpha)^3\sqrt{(\hat p_{34}\cdot
  p_{12})^2-\hat s_{34} s_{12}}\
\delta(\alpha^2s_{12}-2\alpha\, p_3\cdot p_{12}-\hat s_{34})\nn\\
&\times&\left[\frac{d\,p_1}{(2\pi)^3}\,\frac{d\,p_2}{(2\pi)^3}\,\frac{d\,p_3}{(2\pi)^3}
\delta(p_1^2) \delta(p_2^2) \delta(p_3^2)\delta(Q-p_1-p_2-p_3)\right]\ ,
\eeqa
with $\hat p_1=(1+\alpha) p_1$, $\hat p_2=(1+\alpha) p_2$ and $\hat
p_{34}=p_3-\alpha\,p_{12}$.

By integrating over the branching variable $\alpha$ and selecting the physical solution of
the quadratic equation we obtain the final expression
\beq
d\Phi_3(Q;\hat p_1\hat p_2\hat p_{34})=
(1+\alpha_-)^3\sqrt{\frac{\left(\hat p_{34}\cdot p_{12}\right)^2-\hat
    s_{34}s_{12}}{\left(\hat p_{34}\cdot p_{12}\right)^2+\hat
    s_{34}s_{12}}}
\times d\Phi_3(Q;p_1p_2p_3)\ ,
\eeq
with
\beq
\alpha_-=\frac{\hat p_{34}\cdot p_{12}-\sqrt{\left(\hat p_{34}\cdot p_{12}\right)^2+\hat
    s_{34}s_{12}}}{s_{12}};\ \hat p_1=(1+\alpha_-) p_1;\ \hat
p_2=(1+\alpha_-) p_2;\ \hat p_{34}=p_3-\alpha_-\,p_{12}\ .
\eeq

It is now straightforward to generalize this result. We can generate the massless $n$-particle phase space
from a massless $(n-1)$-particle phase space using the FBPS generator.
Specifically, by branching massless particle $j$ we obtain
\beq\label{FBPSgenerator}
d\,\Phi_n(Q;\{\hat p\})=d\,\Phi_{n-1}(Q;\{p\})\times d\,\Phi_{\rm
  fbps}^{[j]}( \{\hat p\}|\{p\})\ ,
\eeq
where\footnote{If we start from the special configuration
  $d\,\Phi(Q,p_1p_2)$ where both momenta are massless, 
we get the solution $\alpha=-\hat s_{23}/s_{12}=-\hat s_{23}/Q^2$ when
branching momentum $p_2$.}
\beqa
d\,\Phi_{\rm fbps}^{[j]}(\{\hat p\}|\{p\})
&=& (2\pi)^3\left(1+\alpha_j\right)^{2n-3}\times
\sqrt{\frac{\left(\hat p_{jn}\cdot Q_j
\right)^2-\,\hat s_{jn} Q_j^2}{\left(p_j\cdot Q_j
\right)^2+\,\hat s_{jn} Q_j^2}}
d \hat s_{jn}\,d\,\Phi_2(p_{jn};\hat p_j\hat p_n)
\nn\\
\hat p_i&=& (1+\alpha_j)\,p_i\ \ (i\neq j) \nn \\
\hat p_{jn}&=&p_j-\alpha_j Q_j=(1+\alpha_j)p_j-\alpha_j Q\nn \\
Q_j&=&\sum_{i\neq j} p_i=Q-p_j\nn \\
\alpha_j&=&
\frac{\left(p_j\cdot Q_j\right)-\sqrt{\left(p_j\cdot Q_j
\right)^2+\hat s_{jn} Q_j^2}}{Q_j^2}\ . 
\eeqa
We introduced the notation $\{p\}=p_1,p_2,\ldots$ for a list of
momenta in order to simplify the notation. The number of momenta in the
list is clear from the context.

By iterating the above FBPS, we can generate a $n$-particle phase space
from a $m$-particle phase space (with at least 1 massless particle) through intermediary
$k$-particle phase spaces.
Furthermore, we can invert the generated $n$-particle phase space by
applying the appropriate clusterings as defined in eq.~\ref{clustermod}.

\begin{figure}[h]
    \centering
    \includegraphics[width=7.5cm,height=5.83cm]{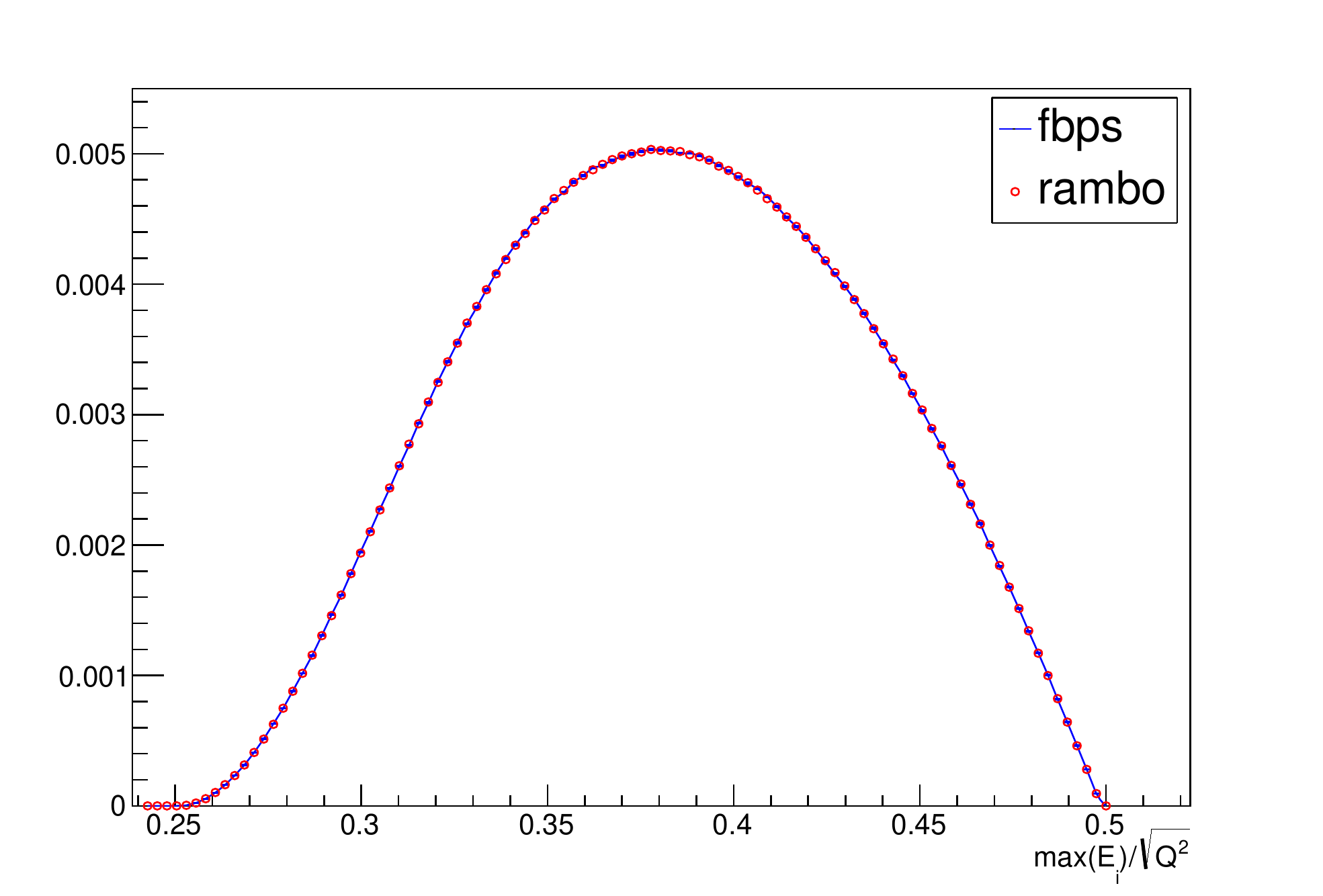}
    \includegraphics[width=7.5cm,height=5.83cm]{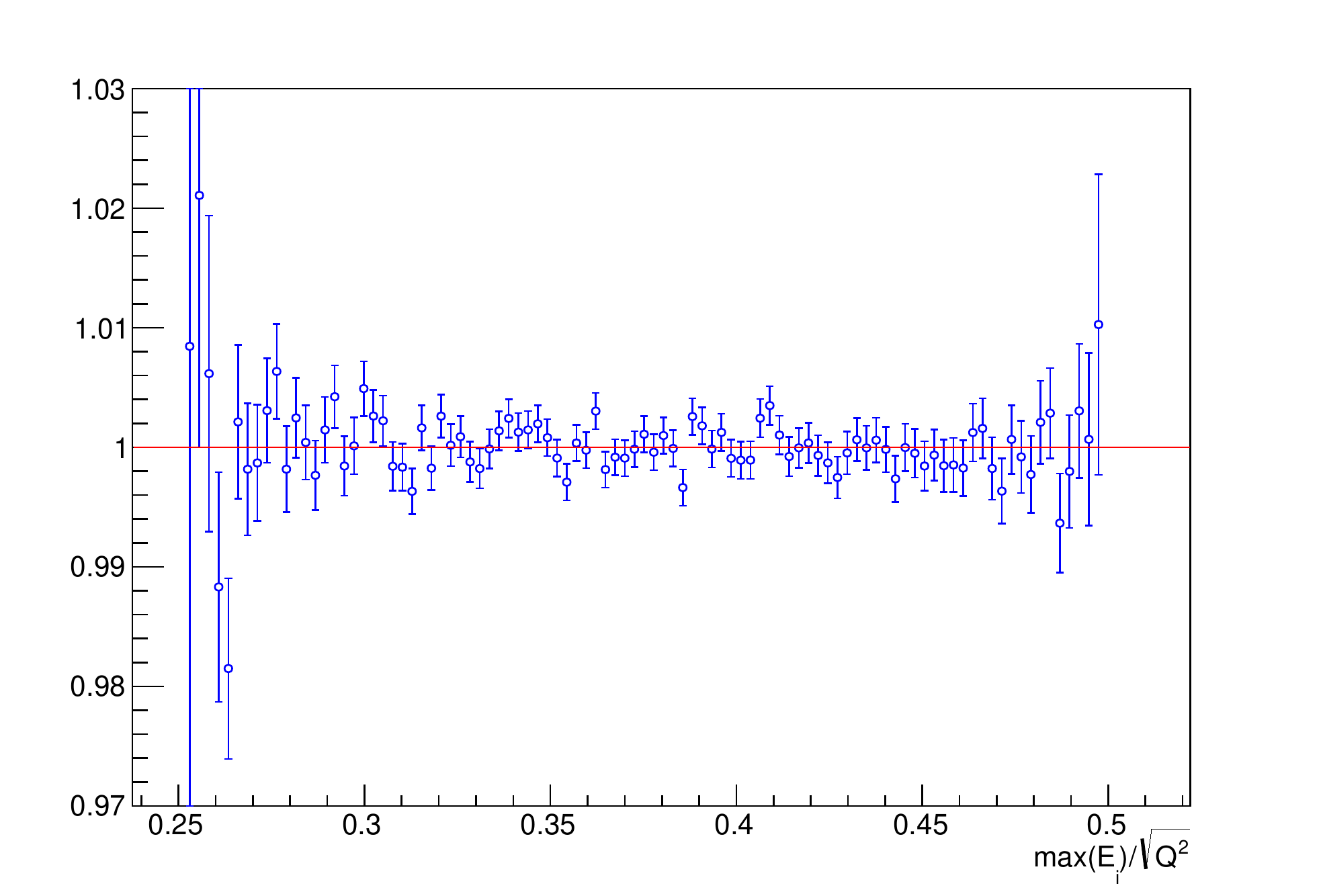}
\caption[]{Comparison of the flat phase space distribution and the
  $3\rightarrow 4$ FBPS generated distribution for the observable
$\max_i(E_i)/\sqrt{Q^2}$ using $10^8$ events. The figure on the right
is the ratio of the two distributions.  Both the 3-particle and
4-particle  flat phase space were generated using RAMBO and
$\sqrt{Q^2}=240$. The results in the left graph were not rescaled for the bin-width.}
\end{figure}
\begin{figure}[h]
    \centering
    \includegraphics[width=7.5cm,height=5.83cm]{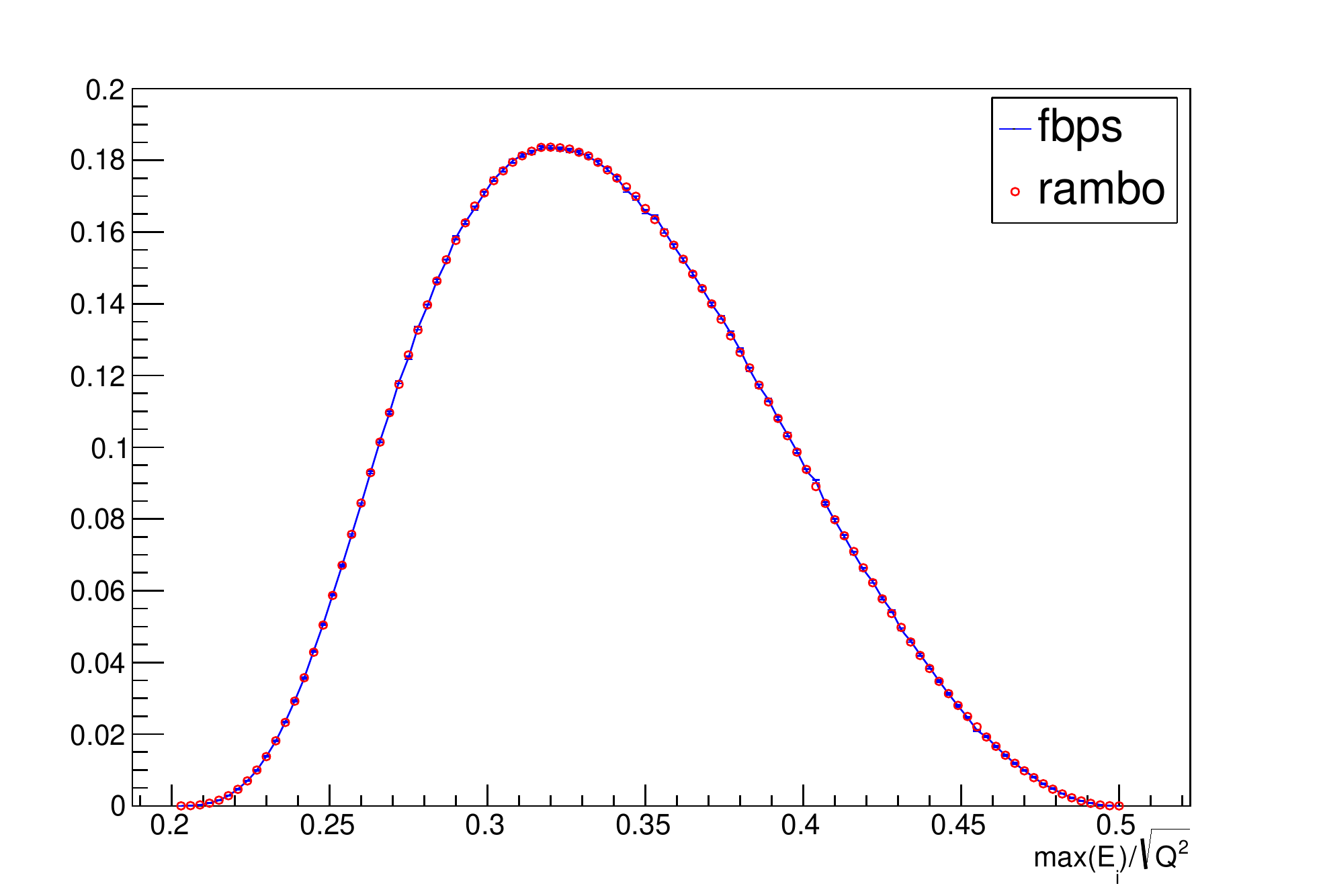}
    \includegraphics[width=7.5cm,height=5.83cm]{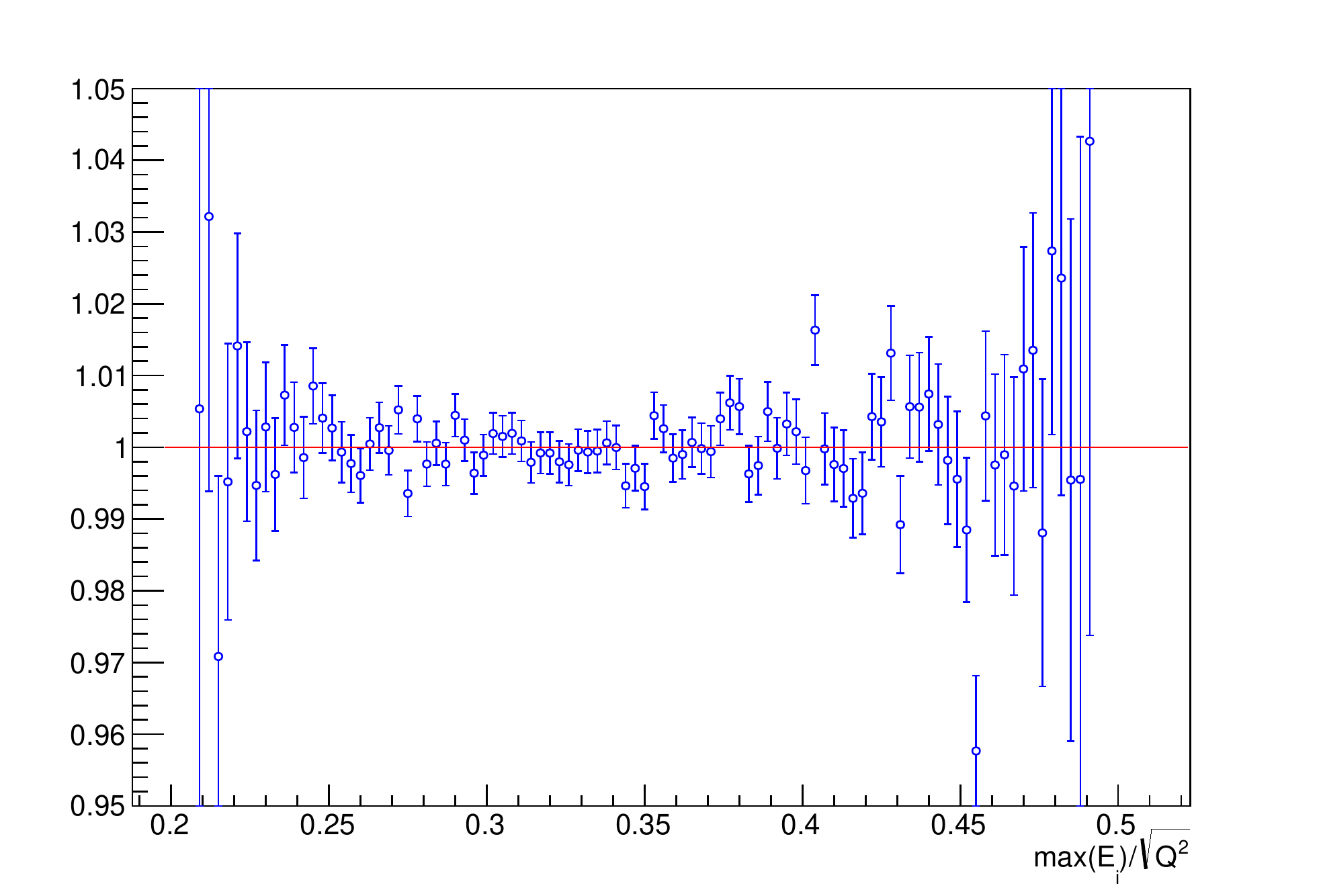}
\caption[]{Comparison of the flat phase space distribution generated
  by and the $3\rightarrow 4\rightarrow 5$ generated FBPS distribution for the observable
$\max_i(E_i)/\sqrt{Q^2}$ using $10^8$ events. The figure on the right
is the ratio of the two distributions.  Both the 3-particle and
5-particle  flat phase space were generated using RAMBO and
$\sqrt{Q^2}=240$. The results in the left graph were not rescaled for the bin-width.}
\end{figure}

 Finally we want to validate the correctness of
 eq.~\ref{FBPSgenerator}. This can be done using a Monte Carlo
 integration of the FBPS generated results versus the results obtained
 using the flat phase space generator RAMBO~\cite{Kleiss:1985gy}.

The first check
is to see whether the FBPS generated phase space has the correct
phase space volume. The volume of phase space is given by 
\beqa
\int d\,\Phi_{n}(Q;\{p^{(n)}\})&=& 
\int
d\,\Phi_{m}(Q;\{p^{(m)}\})\times\left[\prod_{k=1}^{n-m}d\,\Phi_{\rm fbps}^{[j_k]}(\{p^{(m+k)}\}|\{p^{(m+k-1)}\})\right]
\nn\\
&=&\left(\frac{\pi}{2}\right)^{n-1}\frac{1}{(2\pi)^{3n}}\frac{(Q^2)^{n-2}}{(n-1)!(n-2)!}\ .
\eeqa
We numerically verified the thus generated $n$-particle phase space, for
all $3\leq m<n\leq 10$, gives back the correct phase space volume. For
each event generated in the Monte Carlo all $j_k$ were chosen
randomly. The $m$-particle phase space in the comparison was generated
using the flat phase space generator RAMBO.

To further validate the FBPS generator we made the differential cross
section for the observable $\max_i(E_i)/\sqrt{Q^2}$ where $E_i$ is the
energy of momentum $p_i$ for both the $3\rightarrow 4$ FBPS generator in figure 1 and
the $3\rightarrow 5$ FBPS generator in figure 2. The results were compared
against the predictions from RAMBO for the 4- and 5-particle phase
space respectively. As can be seen in both cases the agreement is as
expected, given the $10^8$ generated events used in the Monte Carlo's.

\section {Constructing differential jet cross sections}

Now that we have the basic building block in the form of the FBPS
generator as detailed in eq.~\ref{FBPSgenerator}, we can construct the
event generator to calculate the fully differential jet cross
sections. The event generator will depend on the resolution function
$d_{ij}=d(p_i,p_j)$ of the jet algorithm. The pair of momenta with the
smallest $d_{ij}$ will be clustered by the jet algorithm and it
will keep clustering pairs of momenta until it reaches the condition $\min_{ij}(d_{ij})>d_{\rm cut}$
(exclusive jet cross section) or until a certain multiplicity of
momenta is reached (i.e.. keep clustering until we reach $m$ momenta,
giving us an inclusive $m$-jet cross section).

We will start with eq. ~\ref{FBPSgenerator} and extend it to give us
the generator to calculate the fully differential jet cross sections.
The jet resolution function $d_{ij}$ will do two things for us. It will give which
momentum to branch and set an upper integration boundary on the
variable $\hat s_{jn}$ and decay products of $\hat p_{jn}$.
We can implement the branches by using the resolution function to
partition the $(n-1)$-particle phase space into wedges. Each
wedge is associated with a momentum to branch in order to generate the
$n$-particle phase space.
To do this we define the following partition of one
\beq
1=\sum_{j=1}^{n-1}\theta (\hat d_{jn}=\hat d_{\rm min})\ ,
\eeq
where the $\theta$-function equals one if its argument is true and
zero otherwise. The argument of the $\theta$-function $\hat d_{\rm min}=\min_{ij}(\hat d_{ij})
=\min_{ij}(d(\hat p_i,\hat p_j))$ is true as long as $\hat d_{jn}$ has the
smallest resolution parameter (i.e. it will be the pair which will be
clustered by the jet algorithm).
This partition divides phase space in the desired wedges. In wedge $j$, particle $p_j$
branches to give particles $\{\hat p_j,\hat p_n\}$ in such a manner
that when applying the jet algorithm it will pick this pair to cluster
back to $p_j$ as $\hat d_{jn}$ has the smallest resolution parameter.
Note that this partitioning include $n$-jet final states if we demand
a clustering cutoff when $d_{\rm min}>d_{\rm cut}$. This distinction can be
implemented as 
\beq
1=\sum_{j=1}^{n-1}\theta(\hat d_{jn}=\hat d_{\rm
    min})\Big(\theta(\hat d_{\rm min}<\hat d_{\rm cut})+\theta(\hat d_{\rm
min}>\hat d_{\rm cut})\Big)\ .
\eeq
The $(n-1)$-particle jet exclusive FBPS generator is now given by
multiplying  eq.~\ref{FBPSgenerator}  with the above partition of one
\beq 
d\,\Phi_n(Q;\{\hat p\})=d\,\Phi_{n-1}(Q;\{p\})
\times d\,\Phi_{\rm fbps}^{\rm excl}(\{\hat p\}|\{p\})\,
\eeq
with
\beq
d\,\Phi_{\rm fbps}^{\rm excl}(\{ \hat p\}|\{p\})=
\sum_{j=1}^{n-1}\theta (\hat d_{jn}=\hat d_{\rm min}) \theta(\hat d_{\rm 
min}<\hat d_{\rm cut}) d\,\Phi_{\rm fbps}^{[j]}(\{\hat p\}|\{p\})\ .
\eeq
This FBPS generator will produce all \brem radiation momenta for a
fixed $(n-1)$-jet configuration with jet momenta $\{p_i\}$ provided the
modified clustering of eq.~\ref{clustermod} is used. By integrating
over the jet phase space and removing the $\hat d_{\rm cut}$-constraint,
the full $n$-particle phase space is obtained. We can iteratively
apply this FBPS generator to obtain the multiple \brem phase
spaces.

To define the fully differential jet cross section at NNLO using the improved
clustering is now straightforward. Given the $n$-jet configuration
with massless jet momenta $\{p\}$ we get
\beq\label{xnnlo}
\frac{d^{3n}\sigma^{\rm NNLO}_{\rm impr}}{d^3p_1\cdots d^3p_n}=
{\cal A}(\{p\}) 
+
 \int d\,\Phi_{\rm fbps}^{\rm excl}(\{\hat p\}|\{p\}) 
\times\left[{\cal B}(\{\hat p\}) 
+
\int d\,\Phi_{\rm fbps}^{\rm excl}(\{\hat{\hat p}\}|\{\hat p\}) 
\,{\cal C}(\{\hat{\hat p}\}) 
\right]\ ,
\eeq
where
\beqa
{\cal A}(\{p\})&=&{\cal A}(p_1\cdots p_n)=\left|{\cal M}^{(0)}(p_1\cdots p_n)+{\cal M}^{(1)}(p_1\cdots p_n)+{\cal M}^{(2)}(p_1\cdots p_n)\right|^2
\nn \\
{\cal B}(\{\hat p\})&=&{\cal B}(\hat p_1\cdots\hat p_n)=\left|{\cal
    M}^{(0)}(\hat p_1\cdots\hat p_n)+{\cal M}^{(1)}(\hat p_1\cdots\hat p_n)\right|^2
\nn \\
{\cal C}(\{\hat{\hat p}\})&=&{\cal C}(\hat{\hat p}_1\cdots\hat{\hat p}_n)=\left|{\cal
    M}^{(0)}(\hat{\hat p}_1\cdots\hat{\hat p}_n)\right|^2\ ,
\eeqa
are the appropriate set of matrix elements at each parton multiplicty.
Note that $d\,\Phi_{\rm fbps}^{\rm excl}$ is a 3-dimensional integral,
so that the above integral is at most 6-dimensional.

We can still use the FBPS generator and apply other jet algorithms to
its produced events. There should still be an
advantage over the
usual uncorrelated generation of the different phase spaces.
Given the jet algorithm
mapping $\Delta(\{ p\}|\{\hat p\})$ the differential cross section is
given by
\beqa\label{xxnnlo}\lefteqn{
\frac{d^{3n}\sigma^{\rm NNLO}_{\Delta}}{d^3p_1\cdots d^3p_n}={\cal A}(\{p\}) \Delta(\{ p\}|\{p\})} \nn \\
&+&
\int d\,\Phi_{\rm fbps}^{\rm excl}(\{\hat p\}|\{p\}) \Delta(\{ p\}|\{\hat p\}) 
\times\left[{\cal B}(\{\hat p\}) 
+
\int d\,\Phi_{\rm fbps}^{\rm excl}(\{\hat{\hat p}\}|\{\hat p\}) \Delta(\{\hat p\}|\{\hat{\hat p}\})
\,{\cal C}(\{\hat{\hat p}\}) 
\right]\ .\nn\\
\eeqa
However this fully differential cross section is ill-defined and not
infra-red safe. Suppose we choose all the jet momenta
massless, then the \brem events will never contribute because
the \brem events generate either jets with masses or momentum $Q$ is
changed. That is, in this case the virtual
and \brem events do not merge.
To obtain infrared safety in this case we must define an appropriate observable
such that the \brem and virtual contributions are sufficiently
merged, that is
\beq
\frac{d\,\sigma^{\rm NNLO}_\Delta}{d\,{\cal O}}=\int d\,\Phi(Q;\{p\})\,\delta({\cal
    O}-{\cal O}(\{p\}))\frac{d^{3n}\sigma^{\rm
      NNLO}_\Delta}{d^3p_1\cdots d^3p_n}\ .
\eeq
In this sense the improved jet algorithm is superior to the standard
jet algorithms as it extends infra-red safety to the fully exclusive
jet cross section.

We can generalize eq.~\ref{xnnlo} and~\ref{xxnnlo} using a generating functional $\Gamma$
\beq
\frac{d^{3n}\sigma_{\rm impr}}{d^3p_1\cdots d^3p_n}={\cal
  A}(\{p\})\times\Gamma_{\rm impr}(\{p\}) ;\
\frac{d^{3n}\sigma_{\Delta}}{d^3p_1\cdots d^3p_n}={\cal
  A}(\{p\})\times\Gamma_{\Delta}(\{p\})\ ,
\eeq
to generate the $n$-th order fully differential cross section
calculator by expanding $n$ times the $\Gamma$-functional 
\beqa
\Gamma_{\rm impr}(\{p\})&=& 1+\int d\,\Phi^{\rm excl}_{\rm fbps}(\{\hat
p\}|\{p\})\times\left(\frac{{\cal A}(\{\hat p\})}{{\cal A}(\{p\})}\right)\nn \\
\Gamma_{\Delta}(\{p\})&=& 1+\int d\,\Phi^{\rm excl}_{\rm fbps}(\{\hat
p\}|\{p\})\times\Delta (\{\hat p\}|\{p\})\times
\left(\frac{{\cal A}(\{\hat p\})}{{\cal A}(\{p\})}\right)\ .
\eeqa
These generators can readily be implemented into Monte Carlo programs.

\section{Conclusions}

We constructed a correlated \brem parton level event generator which
should lead to faster convergence of Monte Carlo phase space
integrations for higher order jet cross sections at lepton colliders.
The \brem events are generated highly
correlated, enabling better cancellations between real and virtual
contributions. One can use any jet algorithm on the generated events,
however using the augmented
jet algorithm enables us to define the fully
exclusive multi-jet differential cross section. That is, for any given
jet configuration the generator will integrate out all radiation inside
the jets, thereby making them opaque, and combine the matrix element
weights of the different multiplicities unencumbered by any
constraint. In other words we can define a
probability density (or $K$-factor), calculable order by order in perturbation theory,
for every exclusive jet configuration.

The next step is to extend this method to hadron colliders and apply
the above method to NLO and NNLO multi-jet generators at lepton and
hadron colliders.

\section*{Acknowledgments}
We acknowledge useful discussions with Keith Ellis and
John Campbell.  This research is supported by the US DOE under contract
DE-AC02-07CH11359.

\end{document}